# On Statistical Analysis of the Pattern of Evolution of Perceived Emotions Induced by Hindustani Music – Study Based on Listener Responses

Vishal Midya[1], Sneha Chakraborty[1], Srijita Manna[1], Ranjan Sengupta[2]

[1] *Indian Statistical Institute, Delhi, India*
[2]*C.V.Raman Centre for Physics and Music, Jadavpur University, India*

The objective of this study is to find the underlying pattern of how perception of emotions has evolved in India. Here Hindustani Music has been used as a reference frame for tracking the changing perception of emotions. It has been found that different emotions perceived from Hindustani Music form a particular sequential pattern when their corresponding pitch periods are analyzed using the standard deviations (SD) and mean successive squared differences (MSSD). This sequential pattern of emotions coincides with their corresponding sequential pattern of tempos or average number of steady states. On the basis of this result we further found that the range of perception of emotions has diminished significantly these days compared to what it was before. The proportion of responses for the perceived emotions like Anger, Serenity, Romantic and Sorrow has also decreased to a great extent than what it was previously. The proportion of responses for the perceived emotion 'Anxiety' has increased phenomenally. Both standard deviation (SD) and mean successive squared difference (MSSD) are two very good measures in tracking the changing perception of emotions. The overall pattern of the change of perceived emotions has corresponded to the psychological and sociological change of human life.

## INTRODUCTION

The evolution of emotion has been copiously addressed in recent times. Unlike most other stimuli such as smell, taste or facial expression, that evoke emotion, and can be linked with biological or survival value, Music has an intriguing ability to evoke powerful emotions without having such an intrinsic link. Great classical philosophers—Plato, Aristotle, Spinoza, Descartes conceived emotion as responses to certain sorts of events triggering bodily changes and typically motivating characteristic behaviour**.** It is difficult to find a consensus on the definition of emotion. Most researchers would probably agree that emotions are relatively brief and intense reactions to goal-relevant changes in the environment that consist of many subcomponents: cognitive appraisal, subjective feeling, physiological arousal, expression, action tendency, and regulation. It therefore suggests that some part of the brain would be selectively activated. Origin of emotion may be traced back to 200,000 years ago to semi-nomadic hunter–gatherer. It is argued that their way of living, which involved co-operating in such activities as hunting, avoiding predators, finding food, rearing children, and also competing for resources, could be related to the origin of emotion. Most emotions are presumably adapted from different activities. Several of the activities are associated with basic survival problems that most organisms have in common. These problems, in turn, require specific types of adaptive reactions. A number of authors have suggested that such adaptive reactions are the prototypes of emotions as seen in humans.

If various emotions are cognitively differentiable, as is likely to be publicly agreed upon, there should be differences in sites of brain being excited for different emotions. Bower, an eminent researcher further suggested that every emotion is associated with autonomic reactions and expressive behaviours. These expressive behaviours or responses to the same stimuli can vary depending on many factors external to the stimuli, like the mood of a person, memory association of the person to the applied stimuli.

Objects or a sequence of objects elicit feeling through a sequence of psychophysical processes. The sensory organs convert the signals from the objects to neural pulses. These pulse trains are processed in sub-cortical neural structures, which are primarily inherited. These processed signals produce perception in brain. Through the process of learning and experience we cognize the objects or sequence of objects from these perceived signals. Again through the process of learning we learn to associate these with some emotive environment in the past. This ultimately evokes emotion or feeling. The evaluation of emotional appraisals of stimuli may be done by having the person report the emotions they perceive as reaction to the stimuli. This can be done in several different ways such as verbal descriptions, choosing emotional terms from a list, or rating how well several different emotional terms describe the appraisal. The emotional terms used should be limited in number and as unambiguous as possible. It is also possible to represent these terms in vector forms. The splitting of emotion into dimensions is consistent with Bower's network theory of emotion. However, the number of components and the type of components vary between studies.

In India, music has been a subject of aesthetic and intellectual discourse since the times of *Vedas (Samaveda)*. Rasa was examined critically as an essential part of the theory of art by Bharata in Natya Sastra, (200 century BC). The *rasa* is considered as a state of enhanced emotional perception produced by the presence of musical energy. It is perceived as a sentiment, which could be described as an aesthetic experience. Although unique, one can distinguish several flavours according to the emotion that it colours. Several emotional flavours are listed, namely erotic love (*sringara*), pathetic (*karuna*), devotional (*bhakti*),comic (*hasya*), horrific (*bhayanaka*), repugnant (*bibhatsa*), heroic (*vira*), fantastic, furious (*rudra*), peaceful (*shanta*). [Italics represent the corresponding emotion given in the Indian treatises]. The individual feels immersed in that mood to the exclusion of anything else including himself. It may be noted that during the musical experience, the mind experiences conscious joy even in the representation of painful events because of the integration of perceptual, emotional, and cognitive faculties in a more expanded and enhanced auditory perception, completed by the subtle aesthetic of sensing, feeling, understanding and hearing all at the same time. The Eastern approach to emotional aesthetics and intelligence treats *rasa* as a multi-dimensional principle that explains thoroughly the relation between a sentiment, a mood, the creative process and its transpersonal qualities. This transpersonal domain includes the super-conscious or spiritual state and therefore acts as an interface between individual and collective unconscious states. This transpersonal quality is a germinating power hidden behind aspects of great musical creation that can reveal it, and is able to induce the complete chromatic range of each emotion. In the Vedas the experience of *rasa* is described as a flash of inner consciousness, which appears to whom the knowledge of ideal beauty is innate and intuitive. *Rasa* is not the unique property of the art itself. It unites the art with the creator and the observer in the same state of consciousness, and requires the power of imagination and representation and therefore a kind of intellectual sensibility.

Perception of emotions in human is ever changing and the mechanism or the pattern of change needs to be traced out as it would help us to understand the sociological and psychological changes, the human life is going through. Due to the presence of Hindustani Music for more than 600 years, the changing perception of emotions can be understood by studying the changes in the responses of perceived emotions. For this an extensive study is needed to understand any pattern that emerges from the emotions primarily and then using this fact, we further analyse the evolution of emotions.

## DESCRIPTION OF THE DATA

In Hindustani music, ragas are said to be associated with different rasas (emotions).However, one particular raga is not necessarily associated with one emotion. For the present study we have selected 11 ragas which represent different rasas/emotions residing therein as mentioned in the ancient texts. These emotions are Heroic (Vira), Anger (Rudra), Serenity (Santa), Devotion (Bhakti), Sorrow (Karuna), Romantic (Sringara). In the following table, selected ragas with their corresponding emotions are shown.

TABLE 1

| Name of the Raga | Rasas (Emotions) |
|---|---|
| Adana | Vira |
| Bhairav | Rudra, Santa, Bhakti, Karuna |
| Chayanat | Sringara |
| Darbari Kannada | Santa |
| Hindol | Vira, Rudra |
| Jaijayanti | Sringara |
| Jogiya | Karuna, Sringara, Bhakti |
| Kedar | Santa |
| Mia-ki-Malhar | Karuna |
| Mia-ki-Todi | Bhakti, Sringara, Karuna |
| Shree | Santa |

So from the ancient texts we know which emotions were residing in the ragas 600 years ago. But to know what emotions are evoked now by the same ragas we have used the data extensively from the paper "Alicja Wieczorkowska et al (2010), On Search for Emotion in Hindusthani Vocal music, Springer edited volume entitled 'Advances in Music Information Retrieval' SCI 274, pp.285-304". Some of the important details of the data are discussed now: The aalap part from professional khayal performances are used here. In aalap, characteristics of raga are established. Songs by eminent singers in the ragas mentioned above were selected from the archives of ITC Sangeet Research Academy. From the aalap portions of the ragas, segments of about 30 seconds were randomly taken out for the listening test. Four segments were collected from each aalap. These give 44 sound clips. Along with the emotions mentioned before, the emotion Anxiety has been incorporated in the present study. The emotion Anxiety was absent remarkably in the ancient accounts on the emotions induced by the ragas. The 44 sound clips of ragas were then used for listening test on both Indian and Western listeners. The listeners gave responses of emotions in accordance with what they perceived or what they thought the sequences are evoking as emotions.

## EXPERIMENTAL PROCEDURE

We will now thoroughly study the statistical patterns formed by different emotions induced by these ragas of Hindustani music. The raga clips usually evoke an array of emotions as said before. For the sake of relative ease we have selected the major emotions induced by those clips and we have classified those sequences on the basis of the seven emotions mentioned above, for example under the emotion 'romantic' we have 5 clips and under 'devotion' we have 6 clips etc. For further statistical analysis we obtain the pitch periods corresponding to the selected clips using the software 'Wavesurfer'. The pitch period (in Hz.) of a signal is the fundamental period of the signal, or in other words, the time interval on which the signal repeats itself. The pitch periods are analysed and stored by the software on an average of 10 milliseconds. For statistical computations we have used the software 'Minitab'.

One of the main objectives of the study is to find any pattern in the emotions and we will look for these patterns in the analyzed pitch periods as this is one of the main components of music. A piece of music is a conglomeration of emotion, feature of that particular song and also characteristic of the singer who is rendering that particular piece of music. So in order to minimize the effects of the properties of the song and the characteristics of the singer we stack the sequence of the analyzed pitch periods of a particular clip with the other ones keeping the main emotion evoked by the clips unaltered.

The obtained stacked sequences of pitch periods don't carry any musical significance in the sense of the property of the song and the characteristics of the singer; rather it significantly and prominently contains the effect of emotion. Thus we proceed by analyzing these stacked sequences of pitch periods.

# COMPUTATIONS AND ANALYSES

## COMPUTATION OF TEMPOS

We will start our analysis by calculating the tempos or average number of steady states of the selected clips of ragas. In musical terminology, tempo is the speed or pace of a given piece of music. Tempo is a crucial element of most musical compositions, as it can affect the mood and variation in tempo usually corresponds to variation in emotions evoked from it. Steady state in pitch periods implies that subsequence of pitch periods where all the pitch values lie within a predetermined band around the mean value. The average number of steady states gives a good estimation of the tempo of the piece of music. Hence the values of average number of steady states in any piece of music help us to classify that piece according to the emotions evoked from it. The following table shows the average number of steady states in the clips categorized according to the emotions.

### TABLE 2

| EMOTIONS | TEMPO |
|---|---|
| Anger | 2.0909 |
| Serenity | 1.824625 |
| Romantic | 1.48032 |
| Anxiety | 1.22847 |
| Devotion | 1.209 |
| Heroic | 1.119 |
| Sorrow | 0.596 |

## COMPUTATION OF DESCRIPTIVE STATISTICS

The clips we have taken into consideration don't have the common tonal (the fundamental frequency in the given piece of music). We now calculate the descriptive statistical measures of the sequence of emotions. But we are not going for calculating the means of pitch periods in the sequences as there will be a great bias due to the non uniformity of tonals chosen for the sequences. Change of tonals will affect the mean of the sequences but their variance, skewness, kurtosis and MSSD will not bear any error due to this change in scale. The following table shows the details of the descriptive measures.

### TABLE 3

| VARIABLE | VARIANCE | SKEWNESS | KURTOSIS | MSSD |
|---|---|---|---|---|
| Anger | 7934.28 | -1.11 | 1.06 | 610.91 |
| Serenity | 7197.74 | -0.48 | -0.64 | 418.79 |
| Romantic | 4670.74 | 0.58 | 0.26 | 116.10 |
| Anxiety | 3793.09 | -0.09 | 0.48 | 86.18 |
| Devotion | 3680.85 | -0.26 | 0.64 | 65.29 |
| Heroic | 2561.77 | -0.07 | 0.53 | 53.81 |
| Sorrow | 495.69 | -3.62 | 16.74 | 16.22 |

Comparing Table 2 and Table 3, we get a striking resemblance. The sequential pattern of emotions formed with respect to the average number of steady states remains same when the variance and

MSSD of the emotions are sorted. It can also be noted that the skewness and the kurtosis of the emotion 'sorrow' differ significantly from that of the others.

Now we will test whether the variances of the sequences of the emotions differ significantly from each other.

## TEST FOR NORMALITY AND EQUAL VARIANCES

For conducting the test of equal variance at first we need to find out whether the sequences follow normality. We use the Kolmogorov- Smirnov Test for normality and we get all sequences are non-normal in nature. This result dictates us to use Leven's Test for equal variance (this test is very robust in nature and can be used for any continuous distribution). We set the level of significance at 0.05; the test has been conducted pair-wise taking two sequences of emotions at a time.

The results of the tests show that, except the following pairs of emotions i.e. romantic-devotion, romantic-anxiety and devotion-anxiety all the other emotions have significantly different variances (p-value less than 0.05 in these cases). The following table gives the p-values of the 3 pair wise tests for which the null hypothesis that variances of the two sequences of emotions don't differ significantly, is accepted.

**TABLE 4**

| PAIRS OF EMOTIONS | p-VALUE |
|---|---|
| **Romantic- Devotion** | 0.543 |
| **Romantic-Anxiety** | 0.102 |
| **Devotion-Anxiety** | 0.272 |

It can also be observed that the MSSD of the sequences romantic, devotion and anxiety don't differ significantly. Below is given the box-plot of the three sequences of emotions.

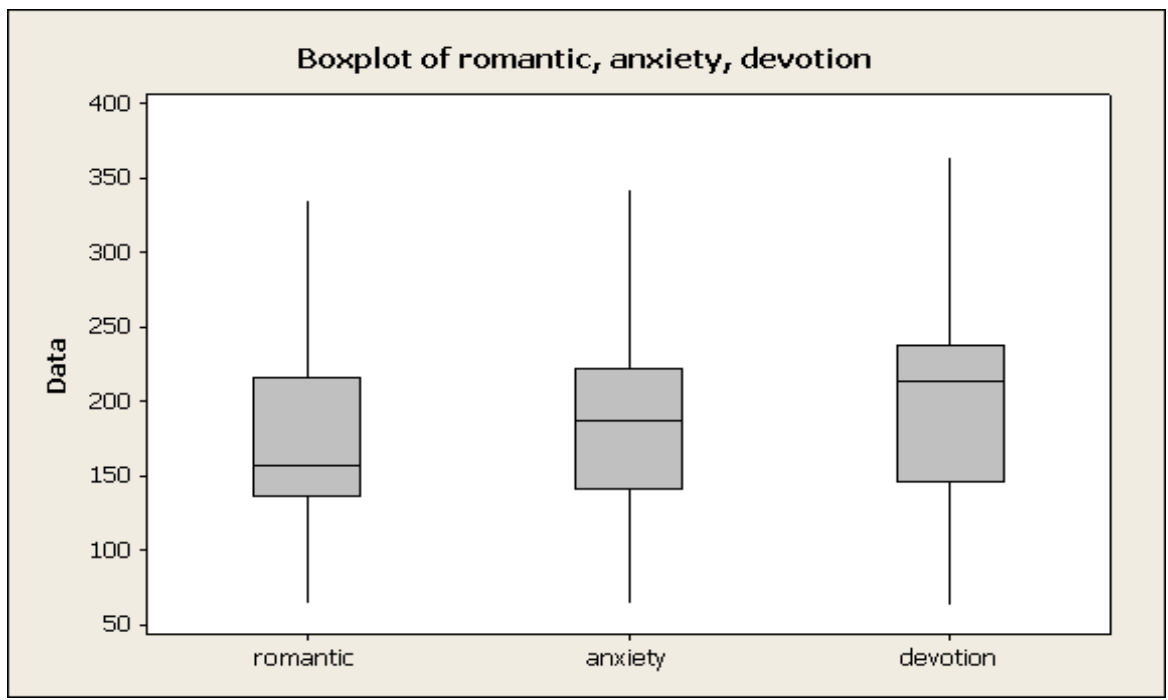

It seems that even if the skewness of the three sequences of emotions differs, the other descriptive features remain same significantly.

## EMERGENCE OF PATTERNS IN THE SEQUENCES OF EMOTIONS

From the above discussion it seems that the sequences of emotions do follow a pattern with respect to their variances and MSSD which is also supported well by their corresponding tempos and we expect that this relative pattern will be maintained even for different sets of ragas sung by different singers in Hindustani music. It should be noted that the absolute values of the variances and the MSSDs of the sequences of emotions are not important in this situation though these values become very crucial in some cases. For example, suppose for a particular set of ragas we know beforehand the values of the variances and the MSSDs of the sequences of emotions (void of feature of the music and characteristics of the singer or instrument) evoked from those. If we now want to find whether a clip of raga evokes a particular emotion, we can then compare the values of the variance and MSSD of the particular emotion with the value of the variance and MSSD of the pitch periods of the clip of raga. If the values don't differ significantly we may conclude that the particular emotion is evoked from the clip or otherwise.

But if the information about the variance and MSSD of the sequences of emotions (void of feature of the music and characteristics of the singer or instrument) of a set of clips of ragas are not known beforehand then we can only sort the clips according to the variances and MSSDs of the pitch periods and expect that these will follow the particular sequential pattern as we have obtained before.

## TEST FOR VARIANCE AND MSSD AS A GOOD MEASURE FOR INDICATING EMOTIONS AND THEIR RELATIVE PATTERNS

We will conduct the test in two ways. At first we will consider 5 ragas namely Hindol, Jaijayanti, Chayanat, Bhairav, Mia-ki-Malhar rendered using sitar by a musician different from the singers of the 44 clips of ragas which had been taken into consideration before for the analysis. The values of the variances and MSSDs of the sequence of emotions for these set of ragas (void of feature of the music and characteristics of the singer or instrument) are not known a priori. The S.D. and the MSSD of the selected ragas are given below.

**TABLE 5**

| RAGAS | S.D. | MSSD |
|---|---|---|
| Hindol(Anger) | 107.92 | 708.99 |
| Jaijayanti (romantic) | 112.90 | 585.80 |
| Chayanat (devotion) | 110.93 | 567.77 |
| Bhairav(anxiety) | 111.43 | 382.39 |
| Mia-ki-Malhar(sorrow) | 102.56 | 399.93 |

We observe that the relative pattern of emotions which had been obtained before is maintained on an average with respect to their variances and MSSDs. The clips of ragas (i.e. the original music clips) do contain the properties of the music, the feature of the instrument and characteristics of the singer along with the emotions. Hence the variances and MSSDs of the pitch periods of these music clips will have some more biases than those sequences of emotions which don't contain any feature of the music or property of the singer or instrument. These combined features have effected in not obtaining the exact relative pattern of emotions though the obtained result is quite satisfactory.

The second test will be conducted using the previously used 44 clips where the variances and MSSDs of the sequences of emotions are known. We frame the test procedure like this: suppose clip 25 used to evoke the emotion serenity 600 years ago but now according to the listeners it evokes the emotion anxiety. Thus the variance and MSSD of the pitch periods of the clip 25 is expected to be different from the variance and MSSD of the sequence of emotion serenity. So by testing the equality of variances by Leven's Test and comparing the MSSDs, we can test how good the measures S.D. and MSSD are.

The test results show that in more than 95% cases, the both S.D. and MSSD are showing accurate results.

So we can now conclude that both these measures enable us to rank the emotions and also in case their values are known apriori we can use both the measures to detect the changing track of emotions.

## RANKING THE EMOTIONS

On the basis of the calculated tempos (or the variance or MSSD) we attach ranks to all the seven emotions shown as below:

**TABLE 6**

| RANK | EMOTIONS | TEMPO | VARIANCE | MSSD |
|---|---|---|---|---|
| 1 | Anger | 2.0909 | 7934.28 | 610.91 |
| 2 | Serenity | 1.824625 | 7197.74 | 418.79 |
| 3 | Romantic | 1.48032 | 4670.74 | 116.10 |
| 4 | Anxiety | 1.22847 | 3793.09 | 86.18 |
| 5 | Devotion | 1.209 | 3680.85 | 65.29 |
| 6 | Heroic | 1.119 | 2561.77 | 53.81 |
| 7 | Sorrow | 0.596 | 495.69 | 16.22 |

Let's proceed to make our intentions clear. We have taken 11 Ragas and the different emotions residing therein (these emotions are the ones that have been reported in the ancient literatures about 600 years ago). We now present the selected ragas with their corresponding emotions (according to the ancient texts) and their respective ranks (according to Table 6). These have been presented in the table below:

**TABLE 7**

| RAGAS | EMOTIONS | RANKS |
|---|---|---|
| **Adana** | Heroic | 6 |
| **Bhairav** | Anger, Serenity, Devotion, Sorrow | 1,2,5,7 |
| **Chayanat** | Romantic | 3 |
| **Darbari Kannada** | Serenity | 2 |
| **Hindol** | Heroic, Anger | 6,1 |
| **Jaijayanti** | Romantic | 3 |
| **Jogiya** | Sorrow, Romantic, Devotion | 7,3,5 |
| **Kedar** | Serenity | 2 |
| **Mia-ki-Malhar** | Sorrow | 7 |
| **Mia-ki-Todi** | Devotion, Romantic, Sorrow | 5,3,7 |
| **Shree** | Serenity | 2 |

Say for example, Adana is associated with the emotion-Heroic. So we have penned rank 6 to it.
Now at present each of the 4 segments of the 11 ragas was randomly played and the listeners' opinions were collected by the listening test. Each of the 4 sequences of a raga evokes an array of emotions but those emotions which most of the listeners have voted for are noted down. The following table thus gives the emotions perceived by the listeners at present for all the 11 ragas.

**TABLE 8**

| RAGAS | EMOTIONS | RANKS |
|---|---|---|
| **Adana** | Anger, Devotion, Devotion, Romantic | 1, 5 ,5 ,3 |
| **Mia ki Malhar** | Sorrow, Sorrow, Anxiety, Heroic | 7 ,7 ,4 ,6 |
| **Mia ki Todi** | Heroic, Devotion, Devotion, Devotion | 6 ,5 ,5 ,5 |
| **Chayanat** | Devotion, Devotion | 5 ,5 |
| **Bhairav** | Anxiety, Anxiety, Anxiety, Anxiety | 4 ,4 ,4 ,4 |
| **Hindol** | Anger, Devotion | 1 ,5 |
| **Jaijayanti** | Heroic, Romantic, Romantic, Romantic | 6 ,3 ,3 ,3 |
| **Jogiya** | Heroic, Heroic, Anxiety | 6 ,6 ,4 |
| **Kedar** | Serenity, Serenity, Serenity, Romantic | 2 ,2 ,2 ,3 |
| **Darbari** | Anxiety, Heroic, Anxiety | 4 ,6 ,4 |
| **Shree** | Devotion, Heroic, Anxiety, Heroic | 5 ,6 ,4 ,6 |

The segments in which the listeners were unable to reach to a conclusion regarding a particular emotion or has reported two different emotions for a particular segment has been neglected to avoid difficulty.

Now we take the previous and present emotions recorded for each raga in two separate columns. For example, Mia ki Todi was previously reported to contain emotions 5, 3 and 7 but now listeners are of the opinion that the 4 segments elicit the emotion 6, 5, 5 and 5. We assume that previously each of the 4 segments contained 5, 3 & 7 simultaneously and since at present the segments are randomized we do not need to think about which segment corresponds to which emotion. So the two columns are constructed in the following manner for Mia ki Todi.

**TABLE 9**

| PREVIOUS RANK OF EMOTIONS | PRESENT RANK OF EMOTIONS |
|---|---|
| 5 | 6 |
| 5 | 5 |
| 5 | 5 |
| 5 | 5 |
| 3 | 6 |
| 3 | 5 |
| 3 | 5 |
| 3 | 5 |
| 7 | 6 |
| 7 | 5 |
| 7 | 5 |
| 7 | 5 |

Like this it is done for all the 11 ragas and we obtain two columns of ranks, namely 'Previous emotions' and 'Present emotions'. By studying the changing patterns of the ranks we can now understand how the perceptions of emotions have undergone the change.

**DESCRIPTIVE MEASURES**

We now obtain the descriptive measures for the two columns of ranks.

**TABLE 10**

| | VARIANCE | IQR | SKEWNESS | KURTOSIS |
|---|---|---|---|---|
| **PREVIOUS EMOTION** | 4.619 | 4.000 | 0.11 | -1.52 |
| **PRESENT EMOTION** | 1.800 | 1.000 | -0.49 | 0.11 |

Clearly a significant variation in the above measures can be observed. We now need to understand how remarkable is these variations as significant change in variance, skewness and kurtosis will imply a changing pattern in the perception of emotions.

**GRAPHICAL REPRESENTATIONS**

From the above results it is very clear that the variability of the ranks have differed significantly.
To have a glimpse of this changing variation we now draw a box-plot which will provide us with a better understanding.

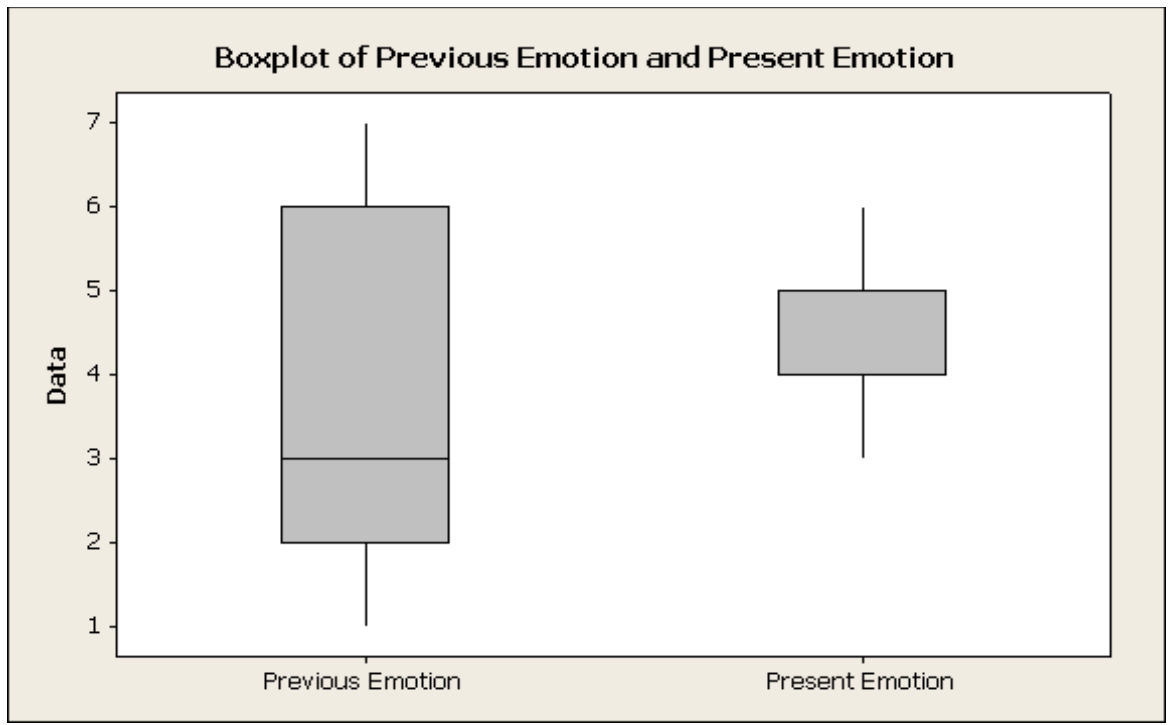

**The box-plot clearly shows that the range of perception of emotions has diminished significantly these days compared to what it was 2200 years ago.**

We also obtain the histograms as follows:

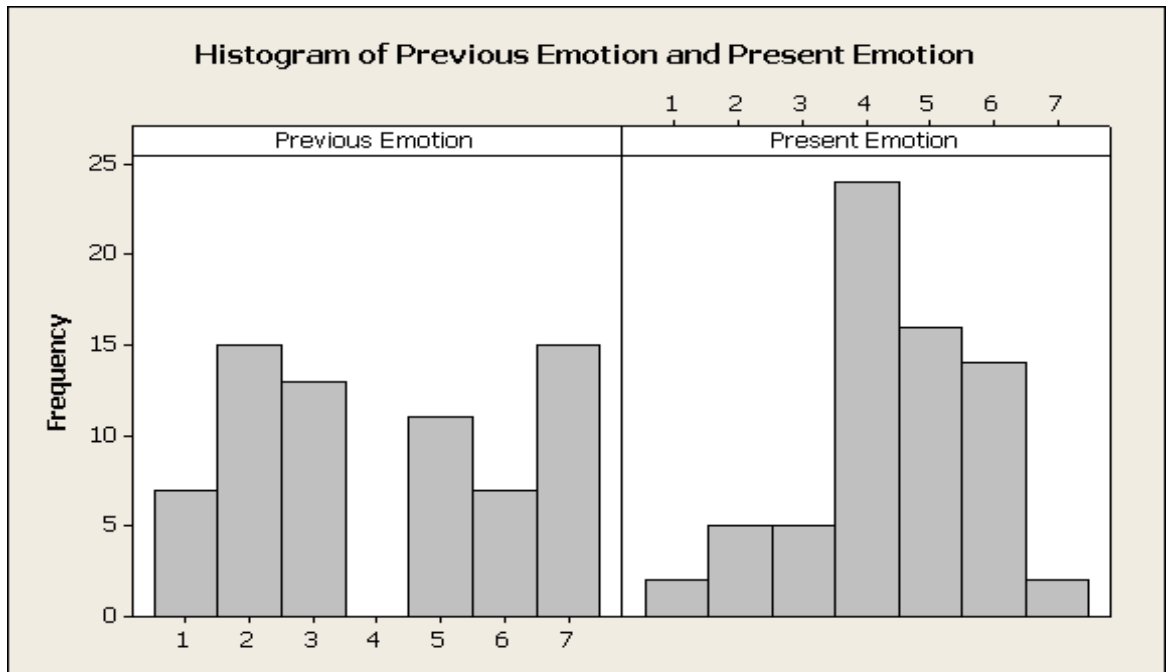

The histograms conveys us a striking picture of the whole data. Previously it is seen that the emotion anxiety (rank 4) has not been perceived at all by the listeners. But at present, in the listening test the listeners have reported "Anxiety" as the most elicited emotion. We can also notice that previously, emotions like Serenity, Romantic and Sorrow was the most apprehended ones while at present these emotions are the less perceived or reported.

## ANALYSES FOR THE CHANGING PERCEPTION OF THE INDIVIDUAL EMOTIONS

Now to understand the scenario for all the emotions individually we proceed in the following manner:

We take into account those segments of the 11 ragas where it has been said to contain the emotion Anger (rank 1) in the past. Now we note down the corresponding emotions that these segments has evoked in the present. For example, Bhairav used to contain the emotion Anger previously but now all of its 4 sequence contain the emotion Anxiety. Again Hindol used to contain the emotion Anger but now it contains Anger and Devotion.

    Bhairav
(1) (Past) ---------------- ⬜(4, 4, 4, 4) (Present)
    Hindol
(1) (Past) ----------------⬜ (1, 5) (Present)

The descriptive measures for each emotion are given below.

**TABLE 11**

|  | MEAN | SD | IQR |
|---|---|---|---|
| **Anger Previous** | 1.000 | 0.000 | 0.0 |
| **Anger Now** | 3.429 | 1.397 | 2.0 |
| **Serenity Previous** | 2.000 | 0.000 | 0.0 |
| **Serenity Now** | 4.000 | 1.363 | 2.0 |
| **Romantic Previous** | 3.000 | 0.000 | 0.0 |
| **Romantic Now** | 4.769 | 1.166 | 2.5 |
| **Devotion Previous** | 5.000 | 0.000 | 0.0 |
| **Devotion Now** | 4.818 | 0.874 | 2.0 |
| **Heroic Previous** | 6.000 | 0.000 | 0.0 |
| **Heroic Now** | 3.571 | 1.618 | 3.0 |
| **Sorrow Previous** | 7.000 | 0.000 | 0.0 |
| **Sorrow Now** | 5.133 | 1.125 | 2.0 |

For the emotion Anger the value of SD and IQR has varied from the previous ones which necessarily imply that the perception of the emotion Anger has been changed significantly. The value of the changed mean of the emotion Anger implies that previously i.e. 600 years ago what we used to perceive as Anger, now we usually perceive that very emotion as Romantic or Anxiety. What we used to perceive as Serenity, now we perceive that emotion as Romantic or Anxiety. Again what we used to perceive as Sorrow, now we perceive that emotion as Devotion or Anxiety. So, in all the cases the perceptions of emotions have gone a vast change.

Let us again look into this mechanism of change in a different way. For each emotion let us calculate the values of the differences of the ranks of present emotion and previous emotion. For example, let us calculate the differences in ranks for the emotion Anger.

TABLE 12

| Anger Previous | Anger Now | Difference in Rank |
|---|---|---|
| 1 | 4 | 3 |
| 1 | 4 | 3 |
| 1 | 4 | 3 |
| 1 | 4 | 3 |
| 1 | 5 | 4 |
| 1 | 1 | 0 |
| 1 | 2 | 1 |

## DESCRIPTIVE MEASURES FOR THE DIFFERENCE IN RANKS FOR ALL THE EMOTIONS

We will now compute the mean and the standard deviations of the difference in ranks of the emotions in order to understand the mechanism of the change of perception of emotions better. Previously there was a stunning absence of the emotion Anxiety that is why neither we can calculate the difference in ranks for Anxiety nor can we compute the descriptive measures. But we can precisely compute the mechanism of the changing perception of the other emotions.

TABLE 13

|  | MEAN | SD |
|---|---|---|
| **Difference in ranks for Anger** | 2.429 | 1.397 |
| **Difference in ranks for Serenity** | 2.000 | 1.363 |
| **Difference in ranks for Romantic** | 1.769 | 1.166 |
| **Difference in ranks for Devotion** | -0.182 | 0.874 |
| **Difference in ranks for Heroic** | -2.429 | 1.618 |
| **Difference in ranks for Sorrow** | -1.867 | 1.125 |

The values of the mean and SD for the emotion Anger implies that what had been commonly perceived as Anger previously is now being reported as Anxiety or Romantic more often. Again previously 600 years ago what we used to perceive as the emotion Sorrow, now most of us perceive that emotion as Heroic or Devotion. Similar changing patterns can also be observed for the other emotions.

These results don't necessarily mean that perceptions of all the emotions when perceived from any music clip of Hindustani music have undergone a huge change rather it shows that in most of the cases the change of perception have been prominent. But in some cases or in the perception of some emotions induced from some particular clips, the change in perception may not have been occurred. That is for example in case of the clip 23; listeners have reported to perceive the emotion Romantic from it and 600 years ago the ancient texts show that it too evoked the same emotion Romantic. The number of cases of this special kind have been limited to a very few. But in most of the cases the present day listeners have differed from what it is written in the ancient texts.

## SD AND MSSD AS TRACKER OF CHANGING PERCEPTION OF EMOTIONS

Table 7 and Table 8 show the detailed list of ragas along with the emotions contained therein as reported in the ancient literatures and as perceived by present day listeners respectively. For both the cases we now obtain the sequences for emotions (void of feature of the music and characteristics of the singer or instrument). From the above analyses, we have observed that what we used to perceive previously has changed vastly in most of the cases. We will now try to find whether SD and MSSD can be used as a tracker of the changing perception of emotions. To make this clear let us look into two examples. Previously what had been commonly perceived as Anger is now being reported as Romantic more often. So the two sequences are more or less comparable to each other. In the present scenario the sequence of emotion Anger has greater SD and MSSD than that of the sequence of Romantic and hence we expect that the sequence of Anger (Previous) should have lower SD and MSSD than that of the sequence of Anger (Present).

Again previously what had been commonly perceived as Sorrow is now being reported as Devotion more often. So the two sequences are more or less comparable to each other. In the present context the sequence of emotion Sorrow has lesser SD and MSSD than that of the sequence of Devotion and hence we expect that the sequence of Sorrow (Previous) should have greater SD and MSSD than that of the sequence of Sorrow (Present). In other words, if we obtain the expected results, then just by analysing the values of SD and MSSD of the sequences of emotions, we can determine how the perception of that particular emotion has changed. Now let us compute the values of the SD and MSSD for the sequence of emotions.

**Table 14**

|  | SD | MSSD |
|---|---|---|
| **Anger Previous** | 68.20 | 382.55 |
| **Anger Now** | 89.07 | 610.91 |
| **Serenity Previous** | 71.27 | 128.01 |
| **Serenity Now** | 84.84 | 421.86 |
| **Romantic Previous** | 54.26 | 88.48 |
| **Romantic Now** | 68.34 | 115.27 |
| **Devotion Previous** | 56.55 | 73.04 |
| **Devotion Now** | 60.77 | 65.27 |
| **Heroic Previous** | 72.87 | 443.20 |
| **Heroic Now** | 50.68 | 55.06 |
| **Sorrow Previous** | 52.36 | 59.04 |

We get a quite remarkable result where the success rate is 100%, i.e. in all the cases SD and MSSD measures have precisely denoted the changing track of the emotions. Let us look at the box plots of the emotions below to better understand the changing track of the emotions and what SD and MSSD have achieved successfully.

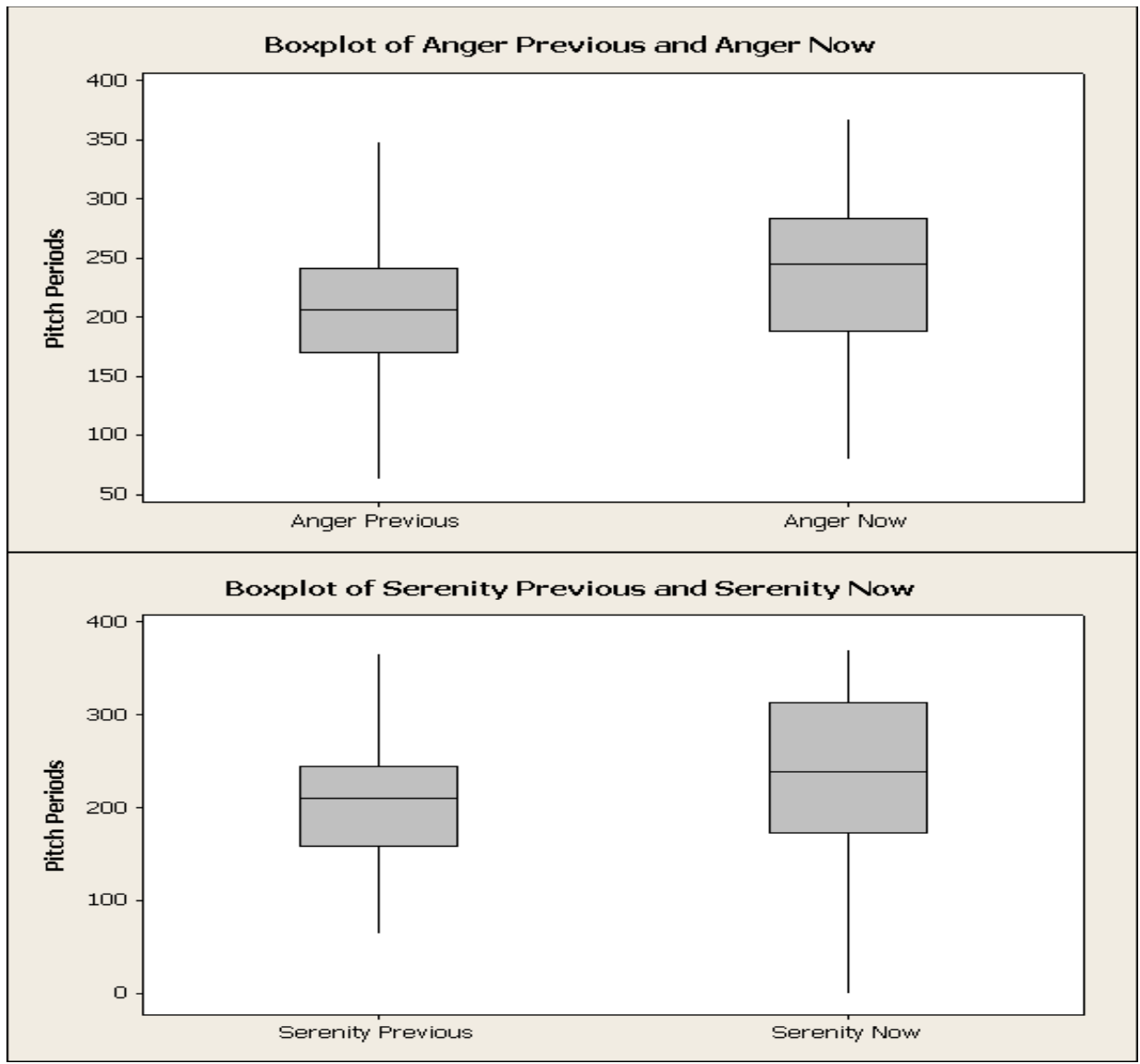

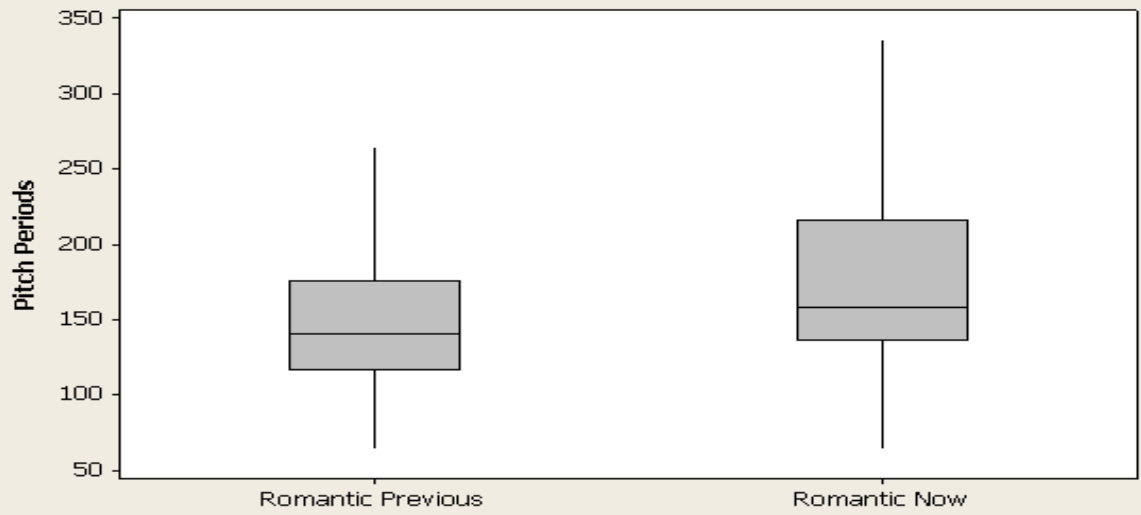
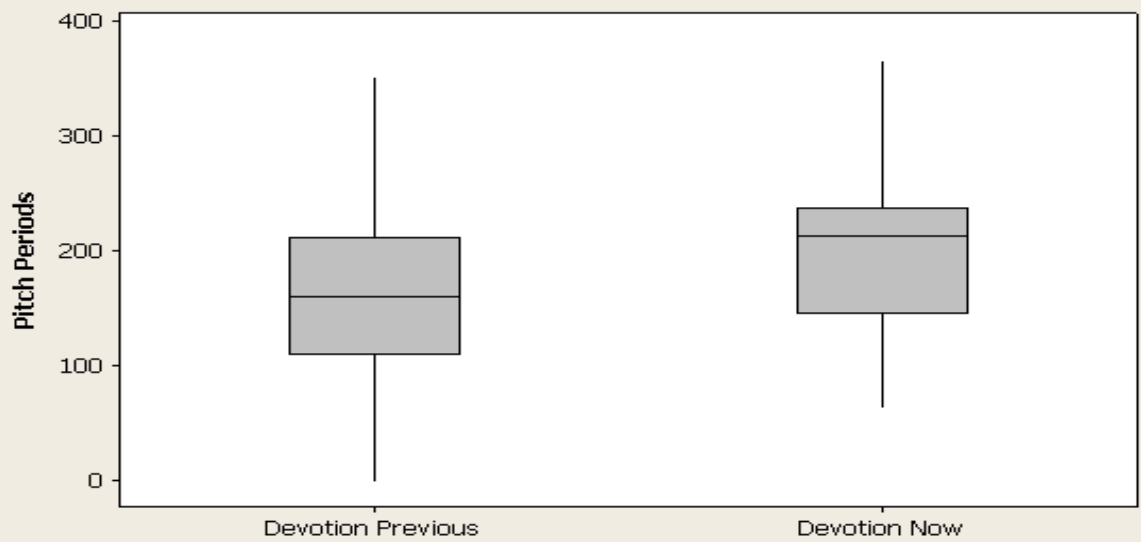

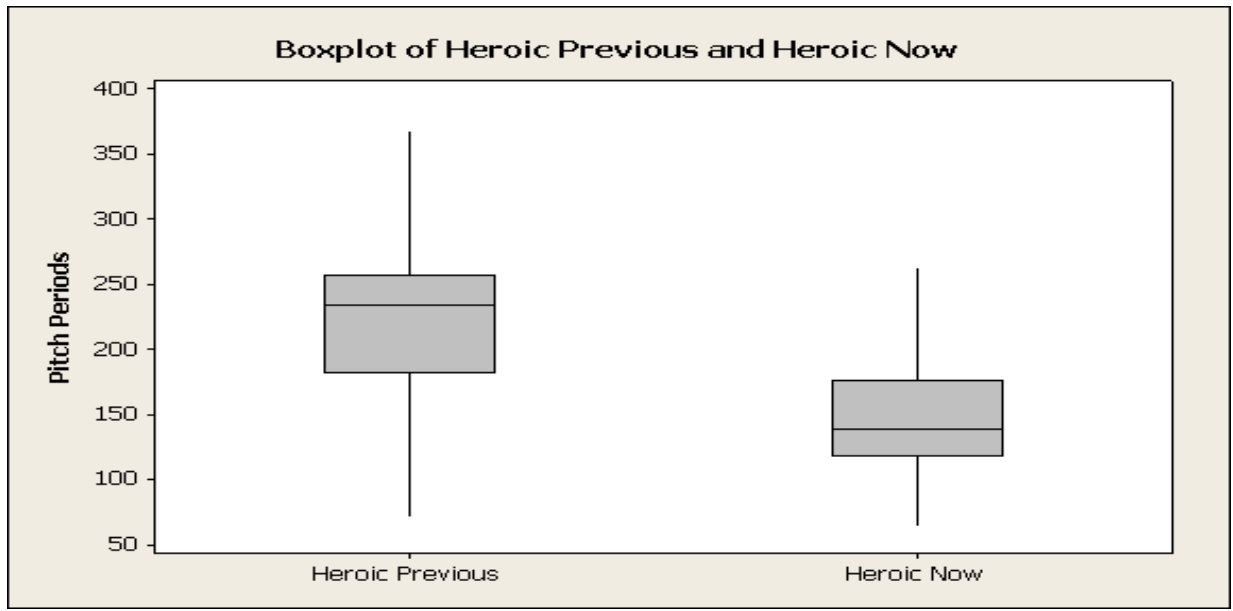
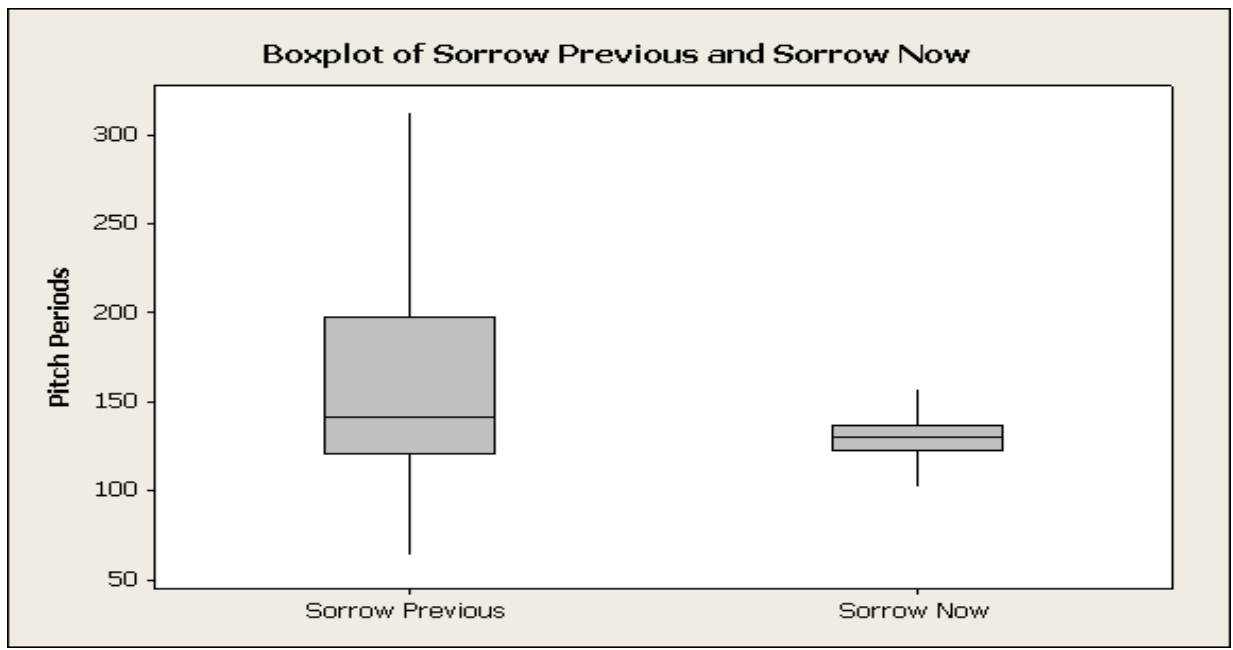

The consecutive box-plots have splendidly shown the changing patterns of emotions. Both SD and MSSD have successfully been able to show this pattern.

## DISCUSSION AND CONCLUSION

What exactly is emotion or how do we perceive emotion is very hard as well as crucial to understand. Studies on this neuropsychological activity are very important in that regard. In this study we have actually been able to find out two very important measures i.e. SD and MSSD which can very efficiently discriminate sequences of emotions and also can form relative sequential pattern among emotions. Surprisingly this sequential pattern among emotions coincides with their corresponding pattern of average number of steady states or tempos. Again it has been established that SD and MSSD measures do serve as a tracker of the changing perceptions of emotions. Undoubtedly validity of these two measures needs to be checked with some more rigour which will then ultimately help us in the sector of artificial intelligence.

The perceptions of emotions have evolved through ages and will probably continue forever. The studies and findings suggest that people at past could sense emotion more distinctively but now they all tend to be mixed into the one most identifiable emotion- Anxiety. The range of perception of emotions has decreased significantly. It should be noted that the pattern of change of perceptions that we found out do support the changing life style and socio-cultural change of human life. The opinions of the listeners may reflect the socio-cultural impact of the contemporary time they belong to. Here also we need to focus more on the reasons of this evolution and how the changing pattern of emotions will ultimately look like.